\documentclass[twocolumn]{aastex62}
\usepackage{indentfirst}
\usepackage{amsmath}
\usepackage[misc]{ifsym}
\begin{document}
\title{\Large \textbf{Constraining the  stochastic gravitational wave from string cosmology with current and future high  frequency detectors }}

\correspondingauthor{X. Fan}
\email{fanxilong@outlook.com}

\author{YUFENG LI}
\affil{Key Laboratory for Computational Astrophysics, 
National Astronomical Observatories, Chinese Academy of Sciences, 
Beijing 100012, China}
\affil{School of Astronomy and Space Science, University of Chinese Academy of Sciences, Beijing 100049, China}

\author{XILONG FAN}
\affil{School of Physics and Technology, Wuhan University, Wuhan 430072, China}

\author{LIJUN GOU}
\affil{Key Laboratory for Computational Astrophysics, 
National Astronomical Observatories, Chinese Academy of Sciences, 
Beijing 100012, China}
\affil{School of Astronomy and Space Science, University of Chinese Academy of Sciences, Beijing 100049, China}

 \begin{abstract}
Pre-Big-Bang models in string cosmology predict a relic background of gravitational wave radiation in the early universe.  The spectrum of this background shows that the energy density rises rapidly with frequency,  which is an interesting target for  high-frequency (i.e., kilohertz) detectors.  In this paper, we discussed the constraining power of multiple configurations of current and future gravitational wave detector (GWD) networks to the stochastic background predicted in string cosmology. The constraining power is jointly determined by the overlap reduction function and the sensitivity curves of multiple detectors. And we further elaborated on the possible contribution of a future Chinese detector and a kilohertz detector to the constraining power of detector network for stochastic background in string cosmology. Our results show that the detectability of the GWD network for the string cosmology gravitational wave background will improve considerably with the joining of a Chinese detector. This is because a Chinese detector (e.g. located at Wuhan ) together with KAGRA, has a better overlap reduction function than the laser interferometer gravitational wave observatory detector pair, and therefore lead to more stringent limits for stochastic background detection. And with ideal overlap reduction function, namely, colocated detectors, a kilohertz sensitivity curve has better performance than previous detectors for stochastic background detection. Finally, the results are compared with the limitations given by the observational constraint of the Big Bang nucleosynthesis bound.
\end{abstract}

\section{Introduction} \label{sec:intro}
The standard cosmological model gives a very good explanation of the present universe, but its universality has been hit by intractable difficulties in explaining the initial singularity~\citep{1993APh.....1..317G}. 
Pre-Big-Bang~(PBB) models in string cosmology may provide a possible explanation for the initial singularity. One of the attractive aspects of string cosmology also lies in the fact that it predicts a quite different gravitational wave background spectrum from that predicted by other cosmological models for the early universe.  Specifically, the spectrum of gravitational waves in string cosmology has rising amplitude with increasing frequency~\citep{1997PhRvD..55.3260A}. This means it falls right into the detection band of ground-based gravitational wave detectors (GWD).  The possible role of ground-based GW detectors for constraining the parameters of string cosmology models was previously discussed in \citet{1999hep.th....7067G}. Moreover, the currently allowed region for the parameters of the relic GW background produced by PBB models, in the light of the most recent observational data, has recently been presented and detailed in \citet{2016JCAP...12..010G}.
We recommend that interested readers find more extensive theoretical pre–Big Bang models and constraining methods in \citet{1999hep.th....7067G} and \citet{2016JCAP...12..010G}. \\
\indent The detectability for string cosmology inevitably depends on the sensitivity and co-response of GWDs. A global network of GW detectors have been proved to play a key role in improving the detection ability to stochastic gravitational wave background in string cosmology~\citep{2008PhLB..663...17F}. Based on the above works, some interesting questions naturally arise, such as, how much would a high-frequency detector (i.e., kilohertz detector) contribute to the stochastic background in string cosmology? Which configuration of GWD network has the best performance and how can we improve the GWD network performance for stochastic background detection in string cosmology? Therefore, in this paper, we tried to answer these questions.\\
\indent In terms of detectability of detectors, a big step has been made in recent years. The gravitational wave astronomy has arrived and been in full swing since The Laser Interferometer Gravitational-Wave Observatory (LIGO) detected the gravitational wave signal from a binary black hole merger on 2015 September 14th. In 2018, the results of gravitational wave searches in the first and second observing runs of the Advanced GWD network were announced in ~\citet{{2018PRL...116...061102L}}. So far, a total of 10 BBH mergers and one binary neutron star (BNS) signal have been identified~\citep{{2018PRL...116...061102L}}. And the results of a search for the isotropic stochastic background using data from Advanced LIGO's second observing run combined with the first observing run have been presented in ~\citet{2019arXiv190302886T}. Then on 2019 April 1st, astronomers and physicists around the world welcomed a long-awaited moment: LIGO in the United States and the Virgo Interferometer in Europe, both of which have significantly improved their detection sensitivity, have officially launched the third run of the year-long gravitational wave experiment (O3). With this upgrade, the probability of LIGO and Virgo finding gravitational wave events will increase significantly, opening a new chapter in our exploration of the universe.\\
\indent Now, KAGRA, a GW detector based on laser interferometry located in Japan, aims to join the third observation run of the advanced LIGO–VIRGO network in late 2019. Once operating along with the existing GW detectors, KAGRA will be helpful in locating GW sources more accurately and determining the source parameters with higher precision, thus improving the detectability of the detector network~\citep{2018arXiv181108079A}.\\
\indent In this context, China is expected to establish a ground-based GWD in the near future. This will naturally lead to a problem: how much contribution will the GWD in China~(CGWD) make to the detector network for stochastic background detection? In the following calculations, the contributions of CGWD to the detector network has been discussed, respectively, in a 2 generation~(2G) era, a 2.5 generation~(2.5G) era, and a 3 generation~(3G) era, so as to estimate the performance of CGWD in different generations, this has certain guiding significance to the construction time of CGWD. Furthermore,  whether CGWD together with KAGRA can operate as an alternative detector when one of the LIGO pairs is offline is also discussed.\\
\indent This paper is arranged as follows: In Section~2, the basic physical scenario of string cosmology and its spectrum has been expounded. In Section~3, we explain how to detect a stochastic background by multiple detectors. In Section~4, the overlap reduction function is introduced; and how to use the overlap reduction function to find the best position and the direction of the detector for stochastic background detection is discussed in Section 5. Finally, in Section 6 and Section~7, the results and a related discussion have been presented.\\

\section{Stochastic background in string cosmology} \label{sec:floats}
\indent In models of string cosmology, the universe passes through two early inflationary stages. The first of these is called the `dilaton-driven' period and the second is the `string' phase. Then after possibly a short dilaton-relaxation era, it came into (radiation then matter dominated) standard cosmology. The two early inflationary stages produce both electromagnetic radiation and stochastic gravitational radiation, then at the end of this stage gravitons decoupled immediately while the electromagnetic radiation went through a complicated history until recombination; this is why we prefer to use gravitational waves to study the early universe~\citep{1995PhLB..361...45B}. \\
\indent The spectrum of gravitational radiation produced in the `dilaton-driven' and `string' phase was discussed in ~\citet{1995PhLB..361...45B}.  In this paper we will use the simplest model. The approximate form of spectrum can be interpreted as follows~\citep{1997gwsd.conf..149B}:\\
\begin{equation}\label{dsp}
\Omega _ { \mathrm { GW } } ( f ) = \left\{ \begin{array} { l l } { \Omega _ { \mathrm { GW } } ^ { \mathrm { S } } \left( f / f _ { \mathrm { S } } \right) ^ { 3 } } & { f < f _ { \mathrm { S } } } \\ { \Omega _ { \mathrm { GW } } ^ { \mathrm { S } } \left( f / f _ { \mathrm { S } } \right) ^ { \beta } } & { f _ { \mathrm { S } } < f < f _ { 1 } } \\ { 0 } & { f _ { 1 } < f } \end{array} \right.
\end{equation}
where $ f_S$ and  $\rm \Omega_{GW}^S$ are frequency and the fractional energy density produced at the end of the dilaton-driven phase respectively. 
And $\rm \beta$ is the logarithmic slope of the spectrum produced in the string phase and is defined by Eq.~\eqref{beta}:
\begin{equation}\label{beta}
\beta = \frac { \log \left[ \Omega _ { \mathrm { GW } } ^ { \max } / \Omega _ { \mathrm { GW } } ^ { \mathrm { S } } \right] } { \log \left[ f _ { 1 } / f _ { \mathrm { S } } \right] }
\end{equation}
where $f_1$ is the maximal frequency above which gravitational radiation is not produced:
\begin{equation}
f _ { 1 } = 1.3 \times 10 ^ { 10 } \mathrm { Hz } \left( \frac { H _ { \mathrm { r } } } { 5 \times 10 ^ { 17 } \mathrm { GeV } } \right) ^ { 1 / 2 }.
\end{equation}
$\rm \Omega_{GW}^{max}$ is the maximum fractional energy density which occurs at frequency $f_1$:
\begin{equation}
\Omega _ { \mathrm { GW } } ^ { \mathrm { max } } = 1 \times 10 ^ { - 7 } \mathrm { h } _ { 100 } ^ { - 2 } \left( \frac { H _ { \mathrm { r } } } { 5 \times 10 ^ { 17 } \mathrm { GeV } } \right) ^ { 2 }.
\end{equation}
$h_{100}$ is a dimensionless parameter for Hubble constant which is generally considered to be in the range of 0.4$\leqslant \rm h _ { 100 } \leqslant 0.85$ by observations~\citep{2008PhLB..663...17F}. $\rm H_r$ is the Hubble factor when the string phase ends and is followed immediately by the thermal radiation dominated phase~\citep{1997PhRvD..55.3882B}. Following~\citet{1997PhRvD..55.3260A}, we assumed $h_{100} = 0.65$ and $H_r = 5 \times 10^{17} GeV$ in this paper.\\
\indent If we make some assumptions about Eq.~\eqref{dsp}, for example, let us set $f_1$ equal to $f_S$, namely $\rm \Omega_{GW}$ vanishes for $f_S<f<f_1$, there will be no stochastic background produced during the string phase of expansion, this is the so-called `Dilaton Only' Case. The spectrum then becomes:
\begin{equation}\label{do1}
\Omega _ { \mathrm { GW } } ( f ) = \left\{ \begin{array} { l l } { \Omega _ { \mathrm { GW } } ^ { \mathrm { S } } \left( f / f _ { \mathrm { S } } \right) ^ { 3 } } & { f < f _ { \mathrm { S } } } \\  { 0 } & { f _ { S } < f } \end{array} \right.
\end{equation}
\\
\indent With the foundation above, how to make use of current GWDs to study stochastic background in string cosmology will be described in the next section.\\

\section{Detecting a stochastic gravitational wave background by multiple detectors} \label{sec:highlight}
\begin{figure}[htbp]
\includegraphics[width=9.5cm]{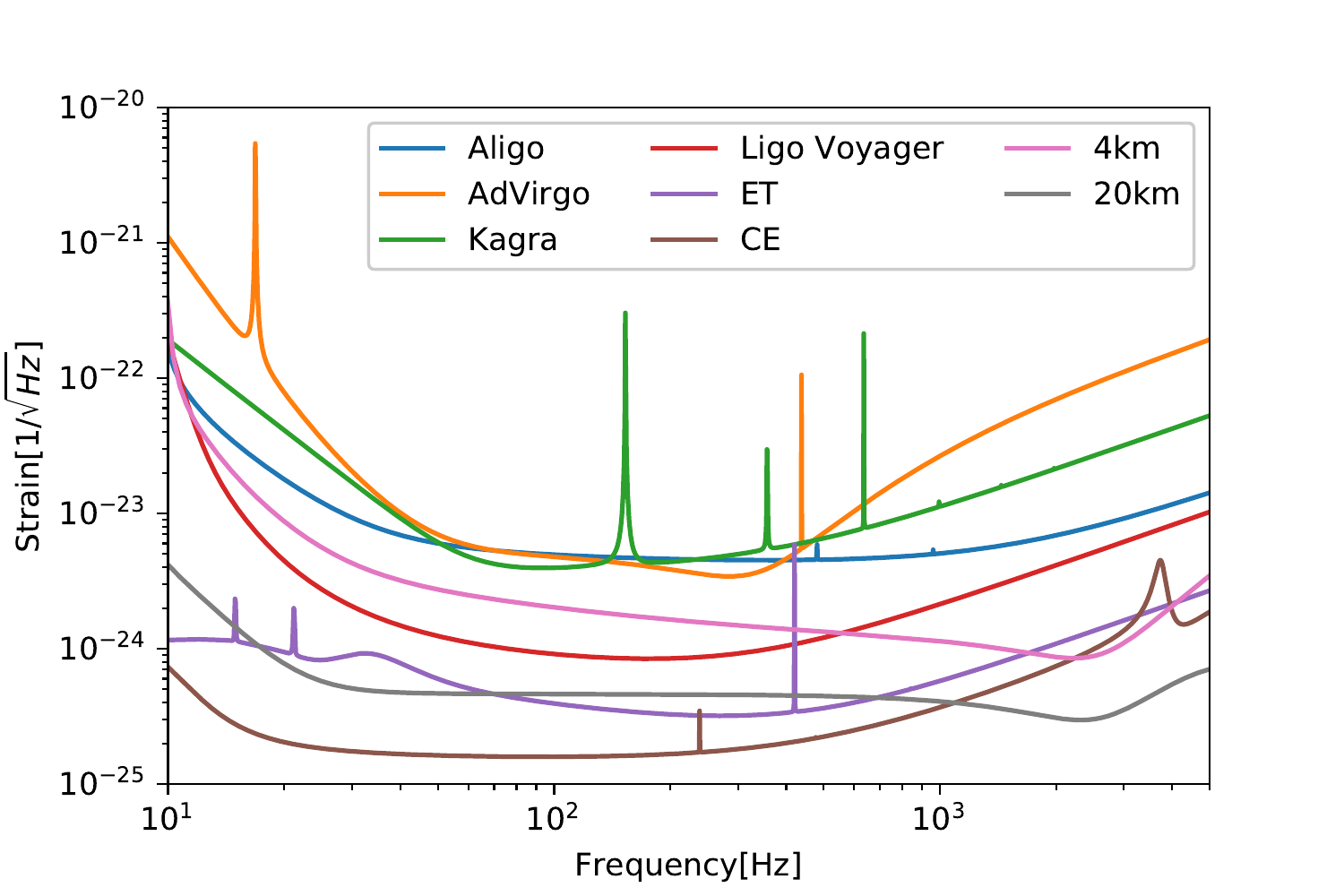}
\caption{Strain curve of multiple detectors.}\label{strain}
\end{figure}

It has been discussed in a series of previous works \citep{1987MNRAS...227..933M, 1992PhRvD..46.5250C, 1993PhRvD..48.2389F, 1999PhRvD..59j2001A} that a network of GWDs can be used to detect a stochastic background of gravitational radiation.  After correlating signals for time $T$~($T = 10^7 s$ = 3 months), the ratio of `signal'~(S) to `noise'~(N) is given by an integral over frequency $f$:
\begin{equation}\label{SN1}
\left( \frac { S } { N } \right) ^ { 2 } = \frac { 9 H _ { 0 } ^ { 4 } } { 50 \pi ^ { 4 } } T \int _ { 0 } ^ { \infty } d f \frac { \gamma ^ { 2 } ( f ) \Omega _ { \mathrm { GW } } ^ { 2 } ( f ) } { f ^ { 6 } P _ { 1 } ( f ) P _ { 2 } ( f ) }
\end{equation}
where the Hubble constant $H_0$ is the rate at which our universe is currently expanding: 
\begin{equation}
H _ { 0 } =  3.2 \times 10 ^ { - 18 }  h  _ { 100 } \frac { 1 } {\rm sec } 
\end{equation}
$P_i(f)$ is the one-side noise power spectral density which describes the instrument noise in the frequency domain.  The one-side noise power spectral density of detectors used in calculations can be found in Figure~\ref{strain}.

\indent Another term that appeared in Eq.~\eqref{SN1} is the overlap reduction function $\gamma(f)$, which will be introduced in detail in Section~\ref{orf}. \\
\indent What has been discussed above is single detector pairs. For multiple pairs of detectors, \citet{2008PhLB..663...17F} has made a detailed comparison of combing multiple detector pairs and directly combine $2N$ detectors, and found that the optimal method is to combine multiple pairs of detectors:
\begin{equation}\label{SNm}
\left( \frac { S } { N } \right) _ { \rm o p t } ^ { 2 } = \sum _ { p a i r } \left( \frac { S } { N } \right) _ { p a i r } ^ { 2 }.
\end{equation}\\
\indent In order to detect a stochastic background with $5\%$ false alarm and $95\%$ detection rate, the total signal-to-noise ratio (S/N) threshold $S/{N_{opt}}$ in Eq.~\eqref{SNm} should be 3.29.  On the whole, we can infer from Eq.~\eqref{SN1} that the detectability of a network of interferometer GWDs to the gravitational wave stochastic background are determined by the noise power spectral density together with the overlap reduction function. On the one hand, if the location and orientation of detector pairs are fixed,  the smaller $P_i(f)$~(the more sensitive the detector) will lead to stronger detectability. On the other hand, if the $P_i(f)$ are fixed, the different location and orientation of detector corresponds to different detectability, that is to say,  the detectability of the detector is much better in certain locations than in others. The same is true in the orientation case. In the next section, we will discuss in detail how the location and arm orientation of the detector affect the S/N.\\

\section{The overlap reduction function}\label{orf}
\begin{figure}[htbp]
\includegraphics[width=9.5cm]{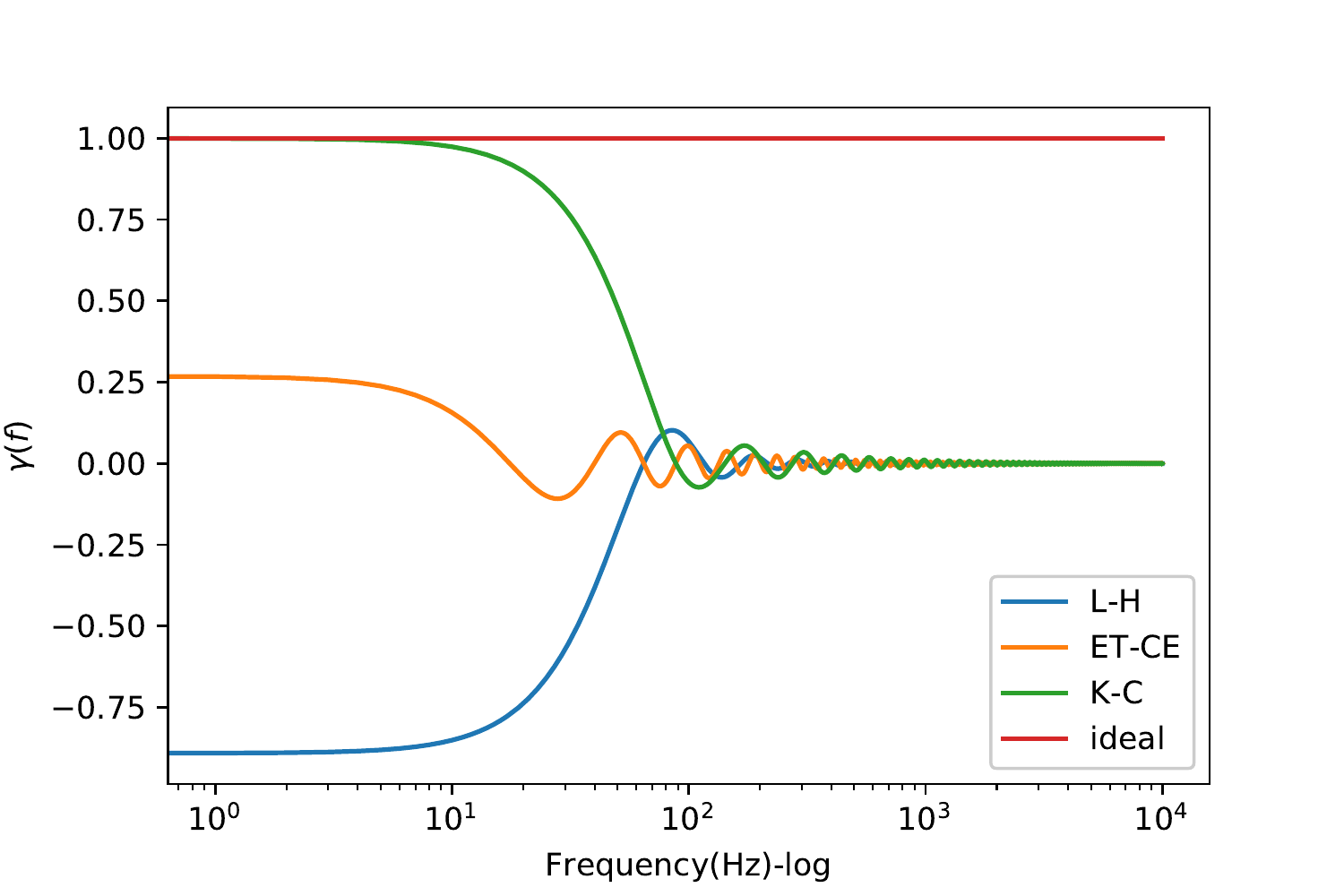}
\caption{Overlap reduction function of various detector pairs. The horizontal axis represents frequency in log. The vertical axis represents the overlap reduction function. Here, `L' is LIGO livingston, `H' is LIGO Hanford, `ET' is Einstein Telescope, `CE' is Cosmic Explorer, `K' is KAGRA, and `C' is a detector located in Wuhan of China which has the same arm orientation as KAGRA. In the `Ideal' case, two detectors are colocated and aligned. }\label{ORF}
\end{figure}

In \citet{1993PhRvD..48.2389F}, the author has developed a detailed analytic formula for the overlap reduction function for the first time. The overlap reduction function is a dimensionless function of frequency $f$, which encodes the relative positions and orientations of a pair of detectors. Explicitly,\\
\begin{equation}\label{gammaf}
\gamma ( f ) : = \frac { 5 } { 8 \pi } \sum _ { A } \int _ { S ^ { 2 } } d \hat { \Omega } e ^ { i 2 \pi f \hat { \Omega } \cdot \Delta \vec { x } / c } F _ { 1 } ^ { A } ( \hat { \Omega } ) F _ { 2 } ^ { A } ( \hat { \Omega } )
\end{equation}
where $\hat { \Omega }$ is a unit vector specifying a direction on the sphere, $\Delta \vec { x } : = \vec { x } _ { 1 } - \vec { x } _ { 2 }$ is the separation vector between the central stations of the two detector sites, and 
\begin{equation}
F _ { i } ^ { A } ( \hat { \Omega } ) : = e _ { a b } ^ { A } ( \hat { \Omega } ) \frac { 1 } { 2 } \left( \hat { X } _ { i } ^ { a } \hat { X } _ { i } ^ { b } - \hat { Y } _ { i } ^ { a } \hat { Y } _ { i } ^ { b } \right)
\end{equation} 
is the $i$th detector's response to a zero frequency, unit amplitude, A represents $+, \times$ polarized gravitational wave, and $\hat { X } _ { i } ^ { a }, \hat { Y } _ { i } ^ { a }$ are unit vectors pointing in the direction of the detector arms.  For coincident, aligned detectors, $\gamma( 0 )$ will be unity. The overlap reduction function of various detector pairs can be found in Figure~\ref{ORF}. Note that the overlap reduction function for the LIGO detector pair is negative as $\rm f \rightarrow 0$. This is because the arm orientations of the two LIGO detectors are not parallel to one another, but are rotated by $90^{\circ}$. In addition, for simplicity, in this paper, we assumed ET has two arms with an included angle of $60^{\circ}$.\\
\indent We can infer from Eq.~\eqref{SN1} that with a larger integral of $\gamma(f)$ over frequency, comes stronger S/N, and stronger S/N corresponds to stronger detectability. Next, we will discuss in detail which factor will, and in what way, affect the integral value of $\gamma$ by Eq.~\eqref{gammaf}. \\
\indent In limit $f \rightarrow 0$, the two detectors become effectively coincident. Because $\gamma(f)$ is an oscillation attenuation function of $f$, the coincidence between detector pairs will decrease when it tends to high frequency. After the overlap reduction function has its first zero, it falls off rapidly at high frequencies. So the most significant part of the integral value of $\gamma(f)$ over frequency is the part before its first zero.
 Naturally we want to figure out which factor will effect the first zero value. In Appendix B of \citet{1993PhRvD..48.2389F}, Flanagan outlines a derivation of a closed-form expression for the overlap reduction function $\gamma( f )$, which is a sum of three spherical Bessel functions. \\
\indent For now, we will just focus on the results of \citet{1993PhRvD..48.2389F}.  First, there will always be frequencies $f$~(as mentioned above, the first zero frequency) for which $\gamma(f)$ vanishes, and correspondingly near which the narrowband sensitivity of the detector pair to the stochastic background is very poor. For detectors that are less than a few thousand kilometers apart, the first zero frequency is at $70 Hz$~(3000 km), irrespective of the detector orientations. This first zero frequency falls off like $1/d$ as the distance increases.  This means that the closer the distance, the larger the integral of $\gamma(f)$ over frequency. Second, $\gamma(f)$ is related to $\beta$, the acute angle between the line joining the two detectors and the plane formed by arms of the detectors. With $\beta$ increasing, $\gamma(f)$ decreases. Finally, if the relative rotation angle between the arm orientation of two detectors is 0, and at the same time, if arms of detectors are parallel to the line joining them, $\gamma(f)$ will be optimal. In a word, if two detectors are close enough, have a parallel arm orientation, and at the same time the arm orientation is parallel to the line joining two detectors, then these two detectors will have an optimal overlap reduction function. Based on these derivations, we can have an idea of how to choose the position and arm orientation of the GWD to achieve the desired overlap reduction function level. In this way, the detection ability of gravitational wave stochastic background can benefit from this by selecting the position and direction of GWDs.  In the next section, we will discuss how to select a better location for the future Chinese detector to have a better performance for stochastic background detection.\\

\section{Location selection for the Chinese detector}\label{sec:highlight}
\begin{figure}[htbp]
\includegraphics[width=9cm]{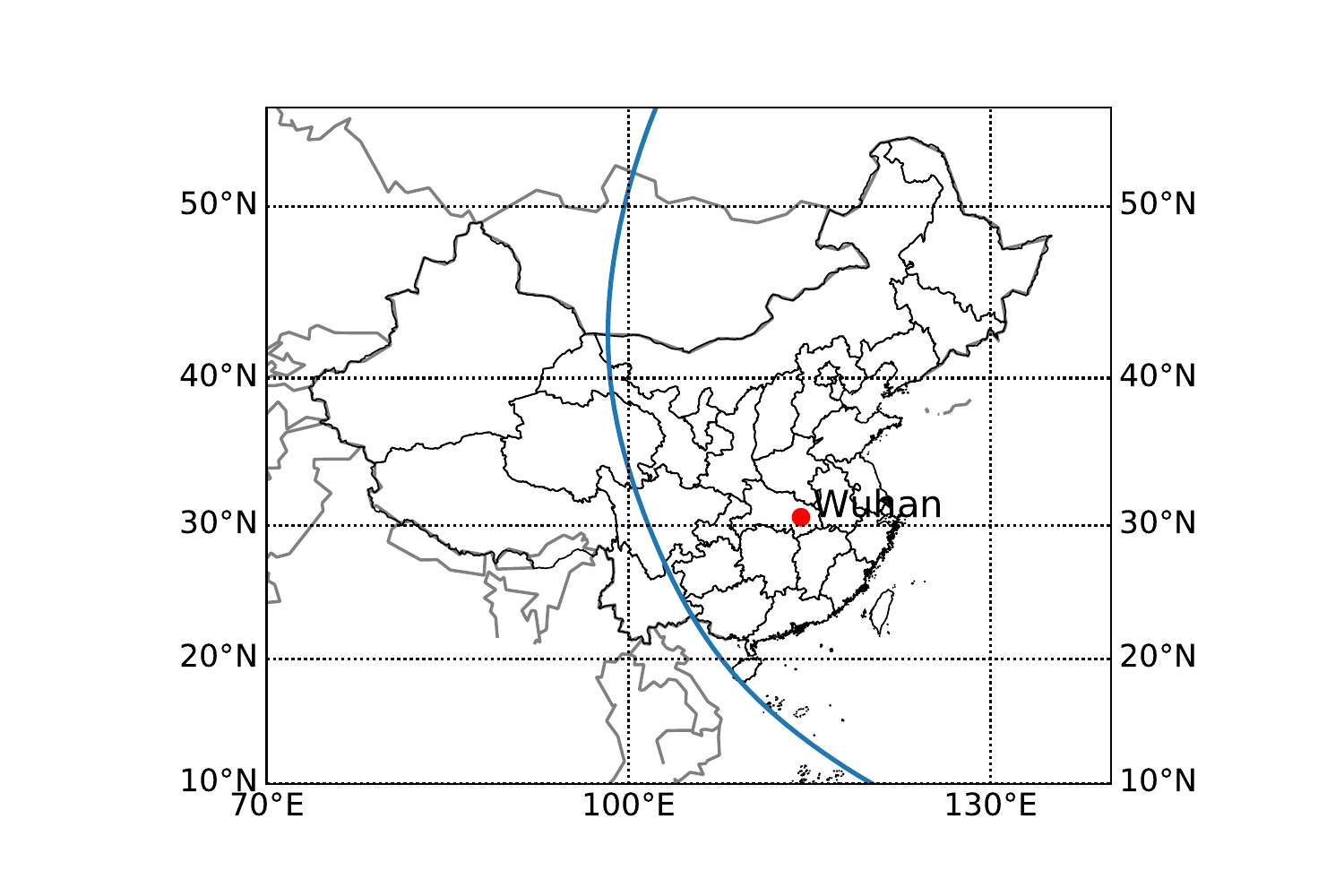}
\caption{Contour of the integral over frequency of square of the overlap reduction function of KAGRA and CGWD located in the range of the figure. The vertical axis represents latitude from north. The horizontal axis represents the longitude from east. The area outlined by the black solid line is China. And a contour line in blue represents the value 14.95 which is equal to the integral of the overlap reduction function of LIGO detector pairs. The area to the right of this blue line represents that the detector located in this area and KAGRA has a larger integral value of the overlap reduction function than LIGO detector pairs.}\label{orf1}
\end{figure}

At present, three detectors are in operation: LIGO Livingston, LIGO Hanford, and Virgo, respectively. In late 2019, KAGRA will join the third observation run of the advanced LIGO-VIRGO network. In this context, we consider where the best place to build a detector in China will be in the future to make it contribute as much as possible to the detector network for stochastic background detection. We know from Eq.~\eqref{SN1} that the detectability of the detector network for stochastic background is positively correlated with the integral over frequency of the square of overlap reduction function and negatively correlated with the integral over frequency of the sensitivity curve. And we have discussed how the overlap reduction function is related to the location and orientation of detectors in Section~\ref{orf}. An obvious conclusion is that the closer the distance, the larger the integral of $\gamma(f)$ over frequency. Providing the location of existing detectors, KAGRA is very close to China, so it occurred to us naturally to consider that the Chinese detector can team with KAGRA in a similar way as LIGO detector pairs. Thus we did some calculations to evaluate the joint detectability of KAGRA and the Chinese detector compared to LIGO detector pairs. Considering the sensitivity of ground detectors, we only used frequencies in the range of [10,1000] Hz. The results are shown in Figure~\ref{orf1}.\\
\indent In Figure~\ref{orf1}, the vertical axis represents the latitude from north, the horizontal axis represents the longitude from east. We can see the outline of China from this figure. This is actually a contour plot. We calculate the integral over frequency of square of the overlap reduction function of KAGRA and CGWD located in every coordinate in this figure, and then outline a blue contour line; it means that CGWD located on this line together with KAGRA has the same integral over frequency of square of the overlap reduction function as that of LIGO detector pairs. Specifically, the integral over frequency of square of the overlap reduction function of LIGO detector pairs is 14.95. Detectors located on the right of this blue contour line will have a larger integral of the overlap reduction function than that of LIGO detector pairs. Based on these results, we simply choose Wuhan as a possible candidate location for later discussions. It can be inferred from Figure~\ref{orf1} that KAGRA together with CGWD in Wuhan has a slightly better overlap reduction function than LIGO detector pairs. For now, CKGO refers to the detector located in Wuhan.\\

\section{Performance of current and future detectors for stochastic background detection}\label{sec:highlight}
\subsection{Plots}
In both the dilaton-only case and the dilaton+string case, we discussed the performance of various detector network configurations for stochastic gravitational wave background detection based on the S/N of 3.29. The difference between the dilaton-only and dilaton+string cases lies in $\Omega_{GW}(f)$. In both cases, the performance of KAGRA-CKGO in the 2G, 2.5G and 3G detector networks is discussed. Given that S/N of 3.29, the relation between $\Omega_{GW}^S$ and $f_S$ constrained by multiple detector pairs  can be calculated according to Eq.~\eqref{SN1}, all results have been presented in following Figure~\ref{res}. Thus the lines in all figures represent S/N=3.29. The area above the line represents the background which can be detected by the detector network with $S/N\ge3.29$. \\
\indent Furthermore, we also plot a restrictive observational constraint of the Big Bang nucleosynthesis (BBN) bound~(solid dodger blue lines) in both cases to see the detective chance for the spectrum. Assuming NO stochastic background is produced during the (more poorly understood) string phase of expansion~\citep{1997PhRvD..55.3260A}, namely in dilaton-only case:\\
\begin{equation}
\Omega _ { \mathrm { GW } } ^ { \mathrm { S } } < 2.1 \times 10 ^ { - 5 } \mathrm { h } _ { 100 } ^ { - 2 }
\end{equation}\\
In the dilaton+string case, the BBN bound follows~\citep{1997PhRvD..55.3260A}:
\begin{equation}
\begin{split}
\int \Omega _ { \mathrm { GW } } ( f ) d \ln f &= \Omega _ { \mathrm { GW } } ^ { \mathrm { S } } \left[ \frac { 1 } { 3 } + \frac { 1 } { \beta } \left( \left( f _ { 1 } / f _ { \mathrm { S } } \right) ^ { \beta } - 1 \right) \right] \\
&< 0.7 \times 10 ^ { - 5 } \mathrm { h } _ { 100 } ^ { - 2 }
\end{split}
\end{equation}
Besides, following~\citet{2008PhLB..663...17F}, we plot a tighter bound~(labeled `$\beta$' bound, dashed dodger blue lines) in the dilaton+string case of all generations. We have plotted all these limits in the following figures.\\
\subsubsection{Sensitivity Curves}
\indent All sensitivity curves we have used in calculations can be found in Figure~\ref{strain}. In Figure~\ref{res}, parentheses in the legend represent used sensitivity curves. In the 2G era, our calculations involve current operational Advanced LIGO, VIRGO,  upcoming KAGRA, and CKGO. For LIGO detector pairs, the sensitivity curve of Advanced LIGO are used. For KAGRA and CKGO, the sensitivity curve of KAGRA is applied. And the sensitivity curve of Advanced VIRGO are used for VIRGO. Multiple detector networks are plotted to estimate the performance of the KAGRA-CKGO pair. Then in the 2.5G era, LIGO will upgrade to LIGO voyager, so the sensitivity curve of LIGO voyager is applied to LIGO detector pairs. Furthermore, the 4km detector in Figure~\ref{strain} is a possible high-frequency future project of China in 2.5G era, so its sensitivity curve is used for KAGRA and CKGO, namely the `K(4km)-C(4km)' case.  In the 3G era, the current proposed ET and CE can work as a team to constrain the stochastic background in string cosmology; we used their designed sensitivity, namely, `ET-CE' case to see the performance of this configuration.  On the one hand, we are looking forward to seeing how many contributions KAGRA-CKGO will make due to their excellent overlap reduction function, so we applied the designed 20km detector sensitivity curve to KAGRA-CKGO, namely the `K(20km)-C(20km)' case,  and the sensitivity curve of CE to KAGRA-CKGO, namely the `K(CE)-C(CE)' case.  On the other hand, in the 3G era, it is possible for China to establish two fully colocated and aligned detectors, one with an arm of 4km, and another of 20km. In this case, the overlap reduction function will be optimal, in other words, it is one at all frequencies. This will directly lead to a better performance in higher frequency(see Sec~\ref{si1} for details). We used the designed sensitivity curve of future 4 and 20km detectors which can be found in Figure~\ref{strain}, to calculate the constraining power of this configuration to stochastic background in string cosmology. This line is labeled as `colocated 4-20km' in Figure~\ref{res}. We also calculated two colocated 20km detectors just for comparison, which is labeled as `colocated 20-20km' in Figure~\ref{res}. High-frequency gravitational wave detectors are a possible future project for China; therefore we assumed that the sensitivity of the detector at high frequencies has improved significantly, specifically, that the sensitivity curve of the 4km detector is $5\times 10^{-25}$ in the frequency range of [1000,10,000] Hz,  while the sensitivity curve of the 20km detector is assumed to be $1\times 10^{-25}$ in the frequency range of [1000,10,000] Hz, with an optimal overlap reduction function to estimate the contribution of high-frequency GWDs to stochastic background detection, this line is labeled as `high-$f$ colocated 4-20km' in Figure~\ref{res}.\\

\subsection{Results}
\begin{figure*}[h]
\begin{minipage}[t]{0.5\textwidth}\
\includegraphics[width=8cm]{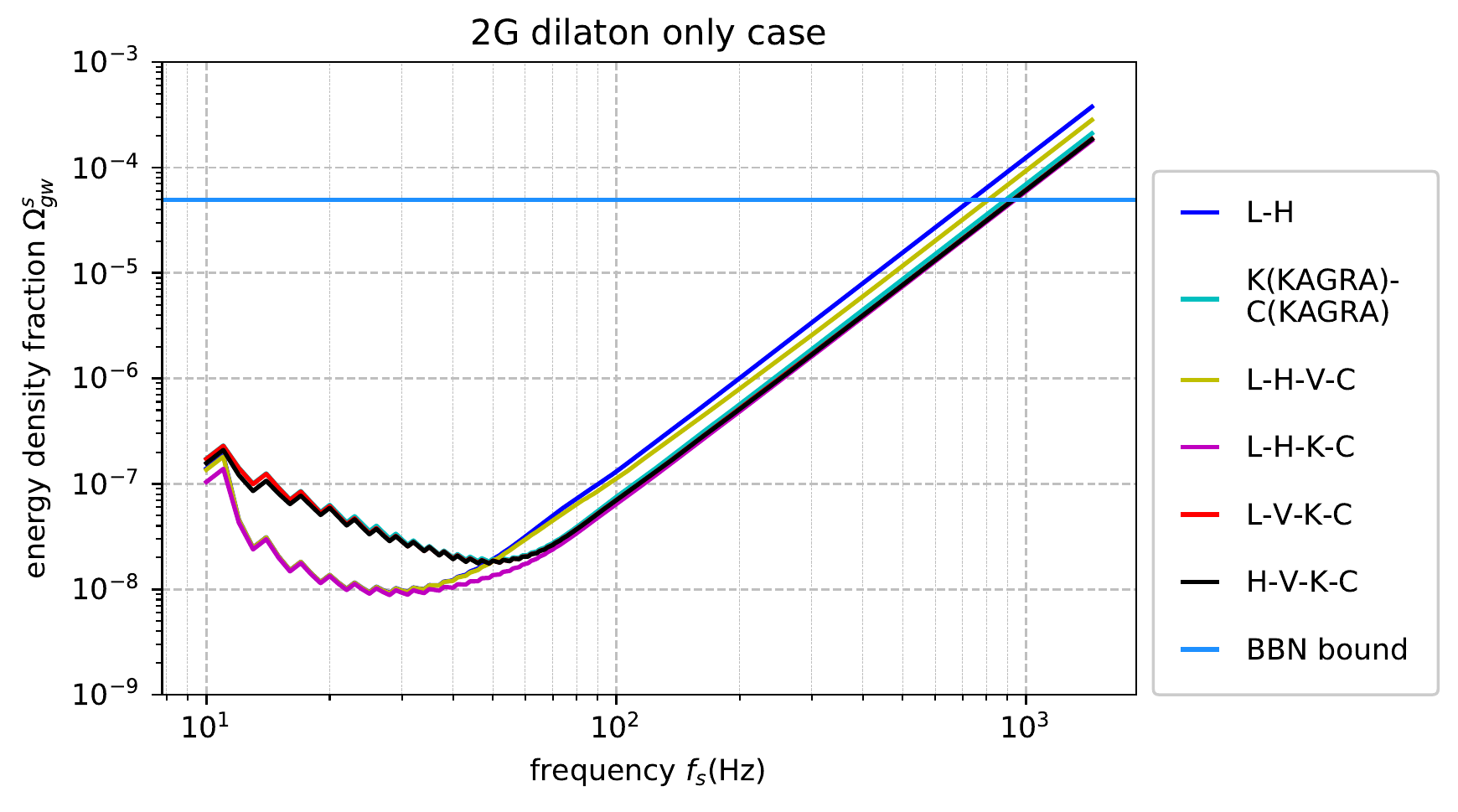}
\end{minipage}
\begin{minipage}[t]{0.5\textwidth}
\includegraphics[width=7.85cm]{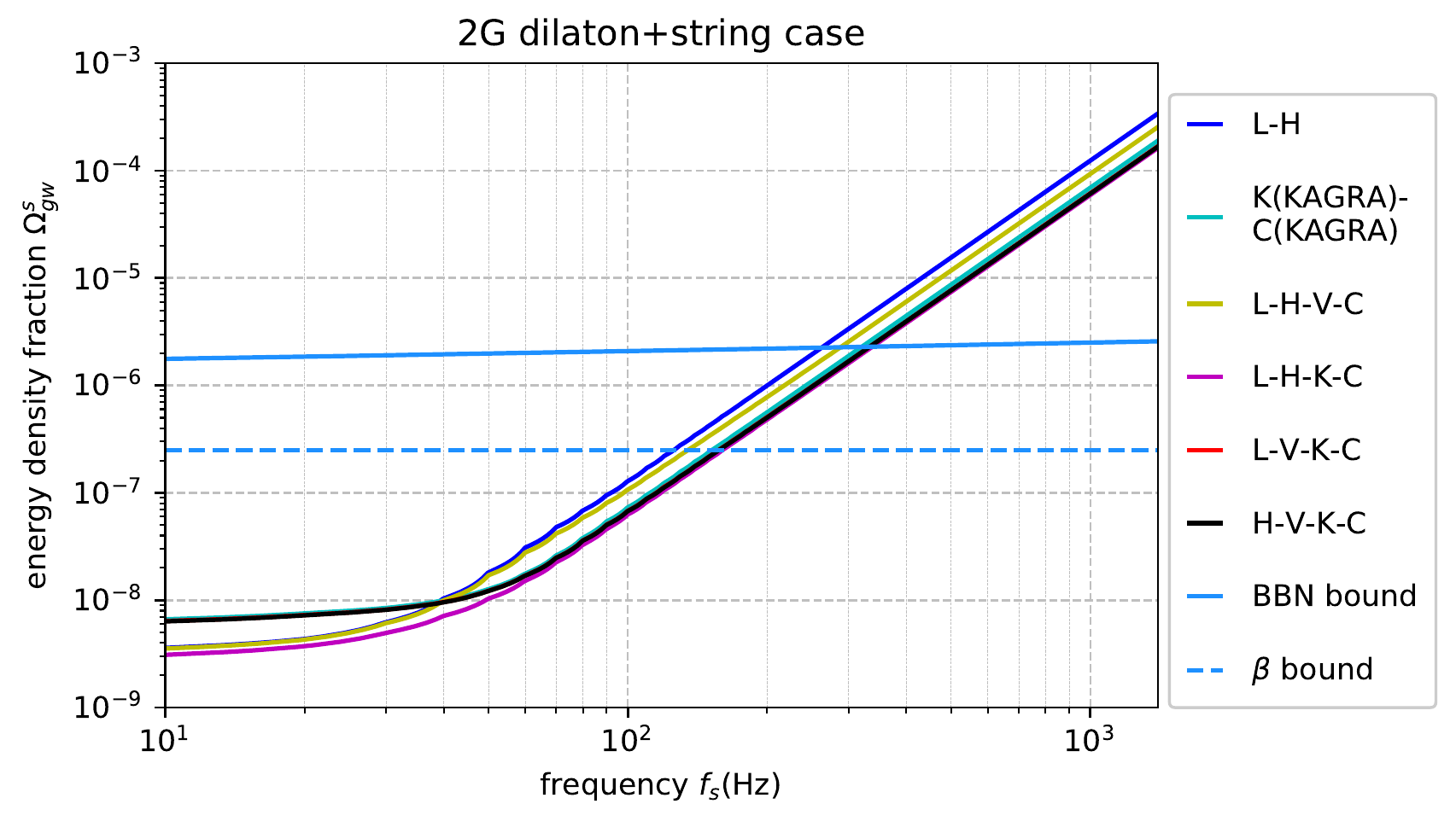}
\end{minipage}
\begin{minipage}[t]{0.5\textwidth}
\includegraphics[width=8.1cm]{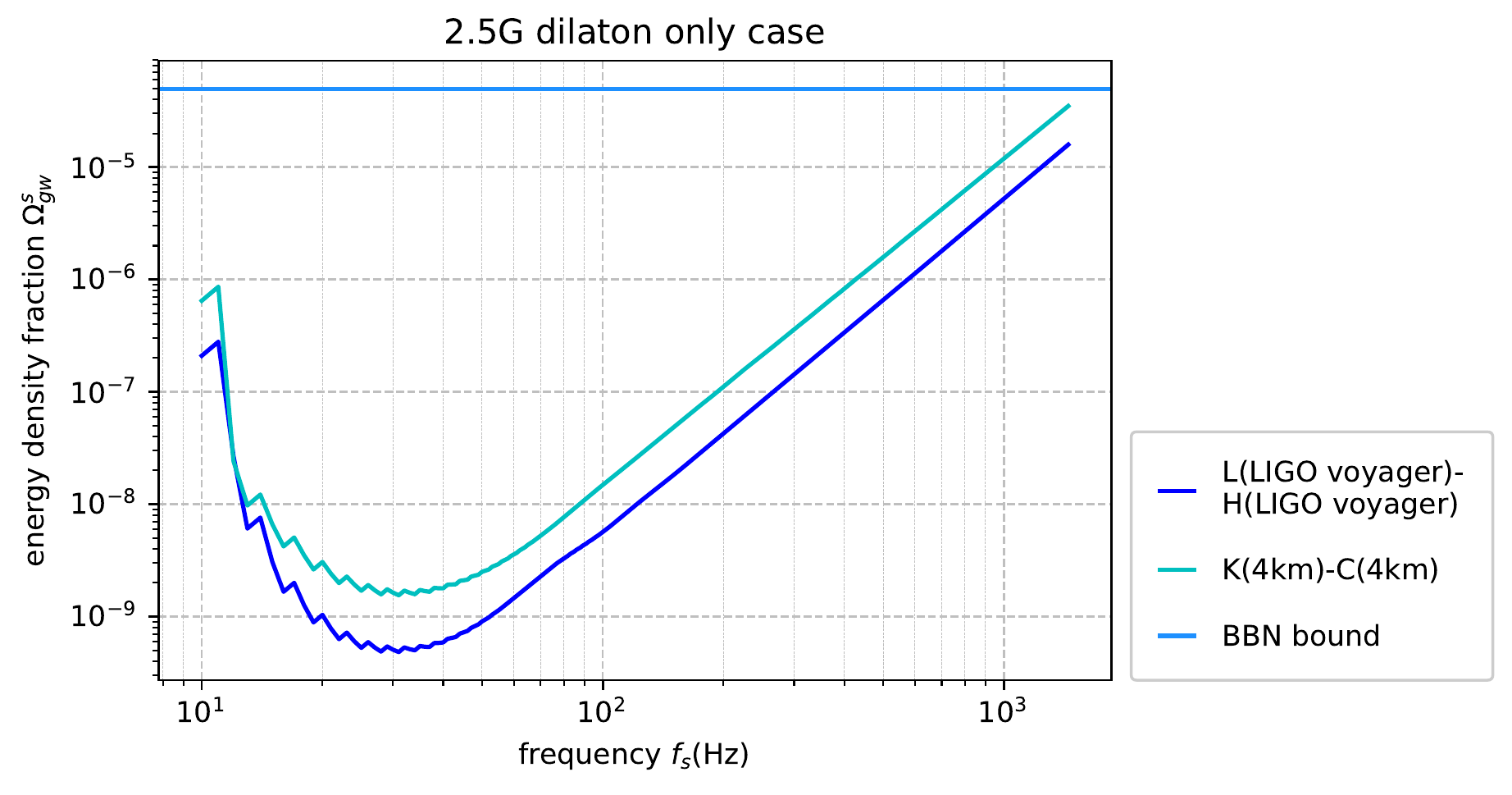}
\end{minipage}
\begin{minipage}[t]{0.5\textwidth}
\includegraphics[width=7.85cm]{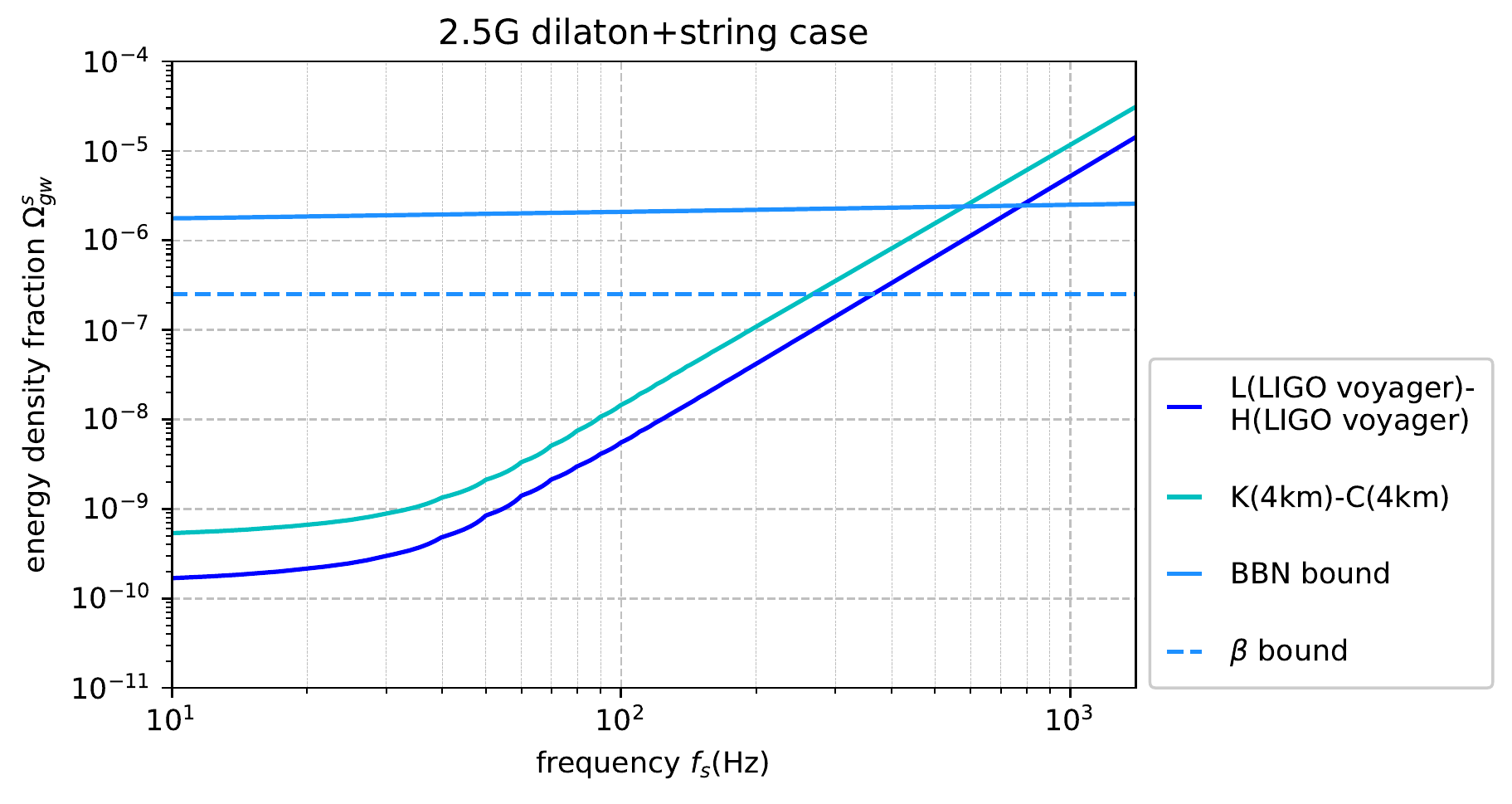}
\end{minipage}
\begin{minipage}[t]{0.5\textwidth}
\includegraphics[width=8.2cm]{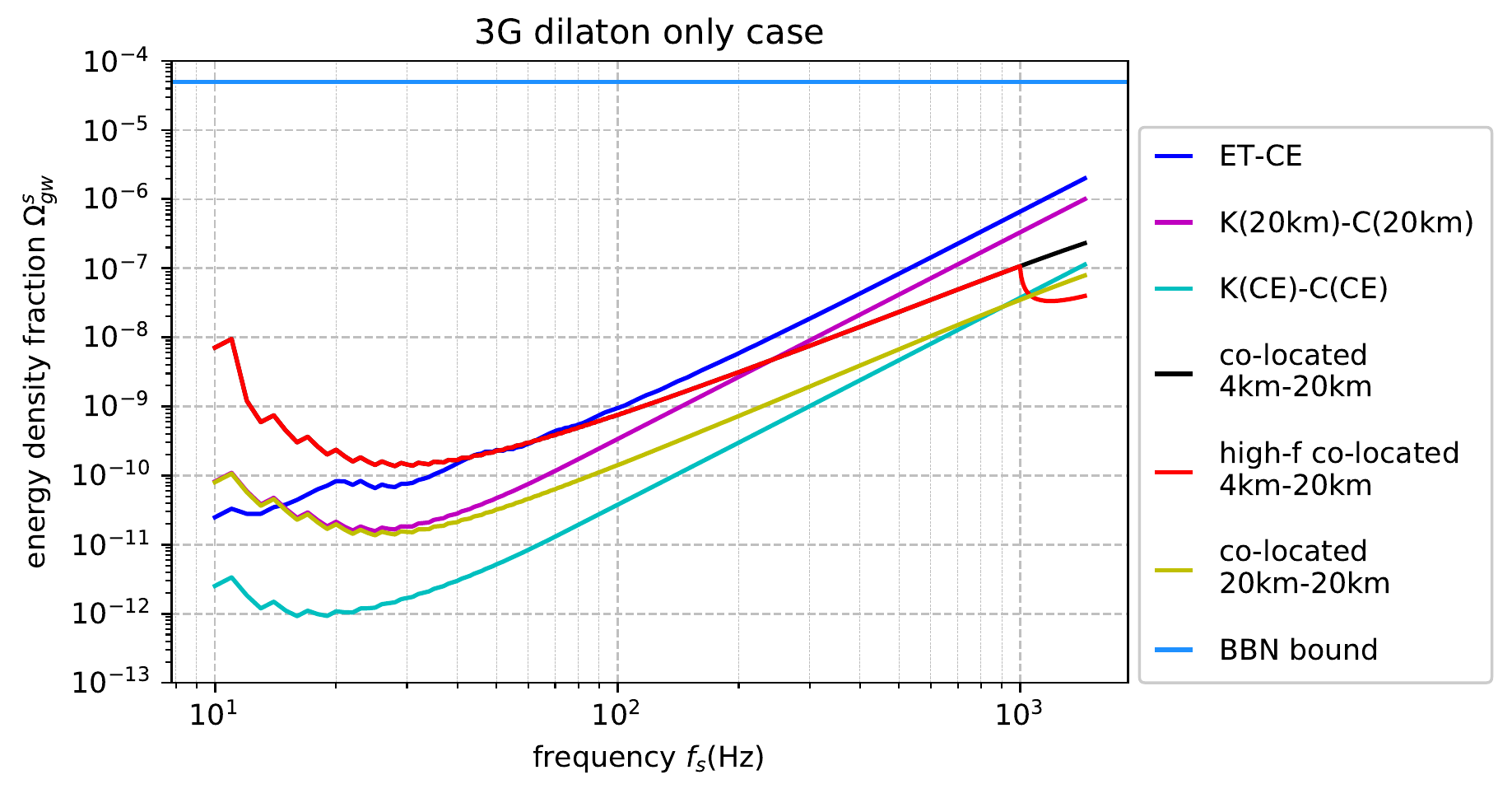}
\end{minipage}
\begin{minipage}[t]{0.5\textwidth}
\includegraphics[width=8.2cm]{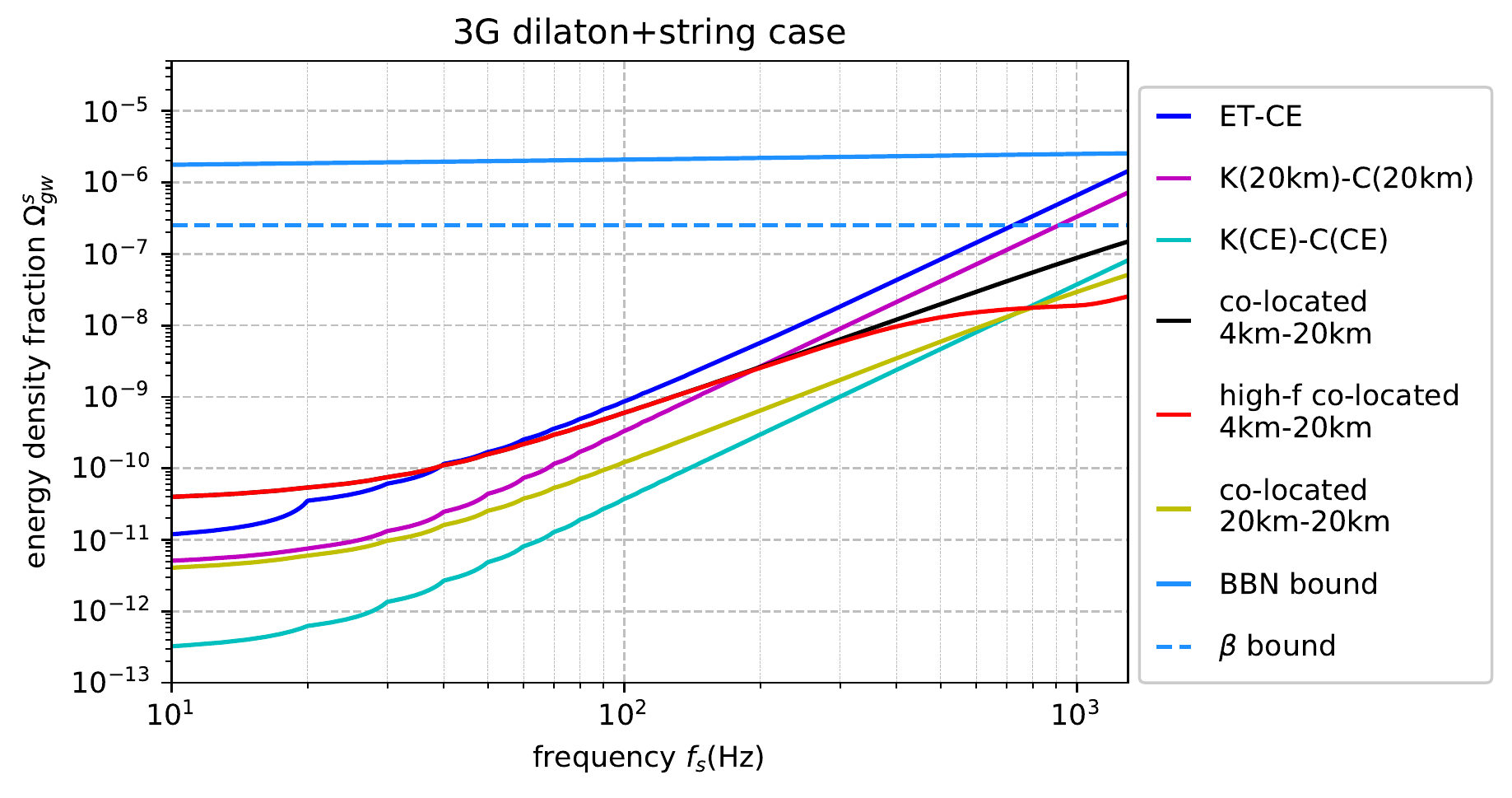}
\end{minipage}
\caption{Lines represent S/N=3.29. `L' is for `LIGO Livingston', `H' is `LIGO Hanford', `V' is VIRGO, `K' is `KAGRA', and `C' is CGWD located at Wuhan. The region above of the curves shows the detectable parameter space of the stochastic gravitational wave background produced in the `dilaton-only' case and `dilaton+string' case by networks of multiple pairs of IFOs with $S/N\ge3.29$.  The BBN bound is shown by solid dodger blue lines in all figures. A tighter bound~(labeled `$\beta$' bound, dashed dodger blue lines) in the dilaton+string case of all generations is also plotted.}\label{res}
\end{figure*}

\indent It can be concluded from Figure~\ref{res} that (1)~For 2G detectors, in a frequency range of higher than~50Hz, the line of KAGRA-CKGO for S/N of 3.29 is lower than that of LIGO pairs. This means that KAGRA-CKGO has much better performance than LIGO detector pairs.  Considering the sensitivity curve of LIGO detector pairs is better than that of KAGRA-CKGO, it indicates that KAGRA-CKGO has a much better overlap reduction function than that of LIGO detector pairs, and the combined effect leads to KAGRA-CKGO making a big contribution to the second generation detector network for stochastic background detection. If one of the LIGO detector pairs is offline, KAGRA-CKGO can work as an alternative of LIGO detector pairs. From the results of multiple detector network configurations, we can see that LIGO detector pairs and KAGRA-CKGO play a major role in the network of L-H-V-K-C. As long as KAGRA-CKGO exists in the network, the whole performance will be better than other configurations. And L-H-K-C has the best performance of all configurations. (2)~For 2.5G detectors, KAGRA-CKGO have weaker constraining power for stochastic background than LIGO detector pairs, this is mainly due to the sensitivity curve applied to LIGO pairs~(LIGO voyager) being better than that of KAGRA and CKGO~(planned 4km detector) below a frequency of 600Hz. In a frequency range of higher than 600Hz, the planned 4km detector has much better sensitivity than LIGO voyager, but it does not show up in the constraining power for stochastic background detection, in detail, this is mainly because the overlap reduction function is infinitesimal and oscillating when it comes to higher frequency. Furthermore, if the sensitivity curve of LIGO voyager is applied to KAGRA and CKGO, the results will turn out to be that KAGRA-CKGO have better performance than LIGO detector pairs. (3)~For 3G detectors, it is evident that KAGRA-CKGO with sensitivity curve of planned 20km detector or CE both have better performance than the ET-CE team. And KAGRA(CE)-CKGO(CE) have unprecedented constraining power for stochastic background detection. In addition, we also discussed the contribution brought by the optimal overlap reduction function. Assuming two colocated and aligned detectors, respectively, with arms of 4 and 20km, the overlap reduction function of these two detectors will be one at all frequencies, just as the red solid line labeled as `ideal' in Figure~\ref{ORF}. Applying sensitivity curves of designed 4 and 20km detectors, the results have been shown in black solid lines labeled `colocated 4km-20km'. We cannot clearly find this black solid line because it overlaps with the red solid line below frequency of 1000Hz.  It can be seen from Figure~\ref{res} that `co-located 4km-20km' configuration has better performance than K(20km)-C(20km) in a frequency range approximately above 250Hz. Considering that the 20km detector is more sensitive than the 4km detector, so an optimal overlap reduction function can lead to better performance in the high-frequency range. Furthermore, the constraining power of high-frequency GWD to stochastic background detection is also discussed by improving the sensitivity curve of both 4 and 20km detectors, namely, `high-f colocated 4km-20km' case stated earlier, the results are labeled by the red solid line in the 3G era in Figure~\ref{res}. It can be easily concluded that with an optimal overlap reduction function, high-frequency GWDs have excellent performance for stochastic background detection in string cosmology when it tends to frequency approximately at 1000Hz. Specifically, in a frequency range of higher than 1000Hz, `high-f colocated 4km-20km' configuration is even better than the `K(CE)-C(CE)' case. Thus, taken together, building a GWD in China and a colocated high-frequency detector pairs would be very profitable for stochastic background detection. The earlier the CGWD is built, the sooner the detection of the stochastic gravitational wave background will benefit from it.\\

\subsection{Sensitivity Integrand}\label{si1}
\begin{figure*}[htbp]
\begin{minipage}[t]{0.5\textwidth}\
\includegraphics[width=8.6cm]{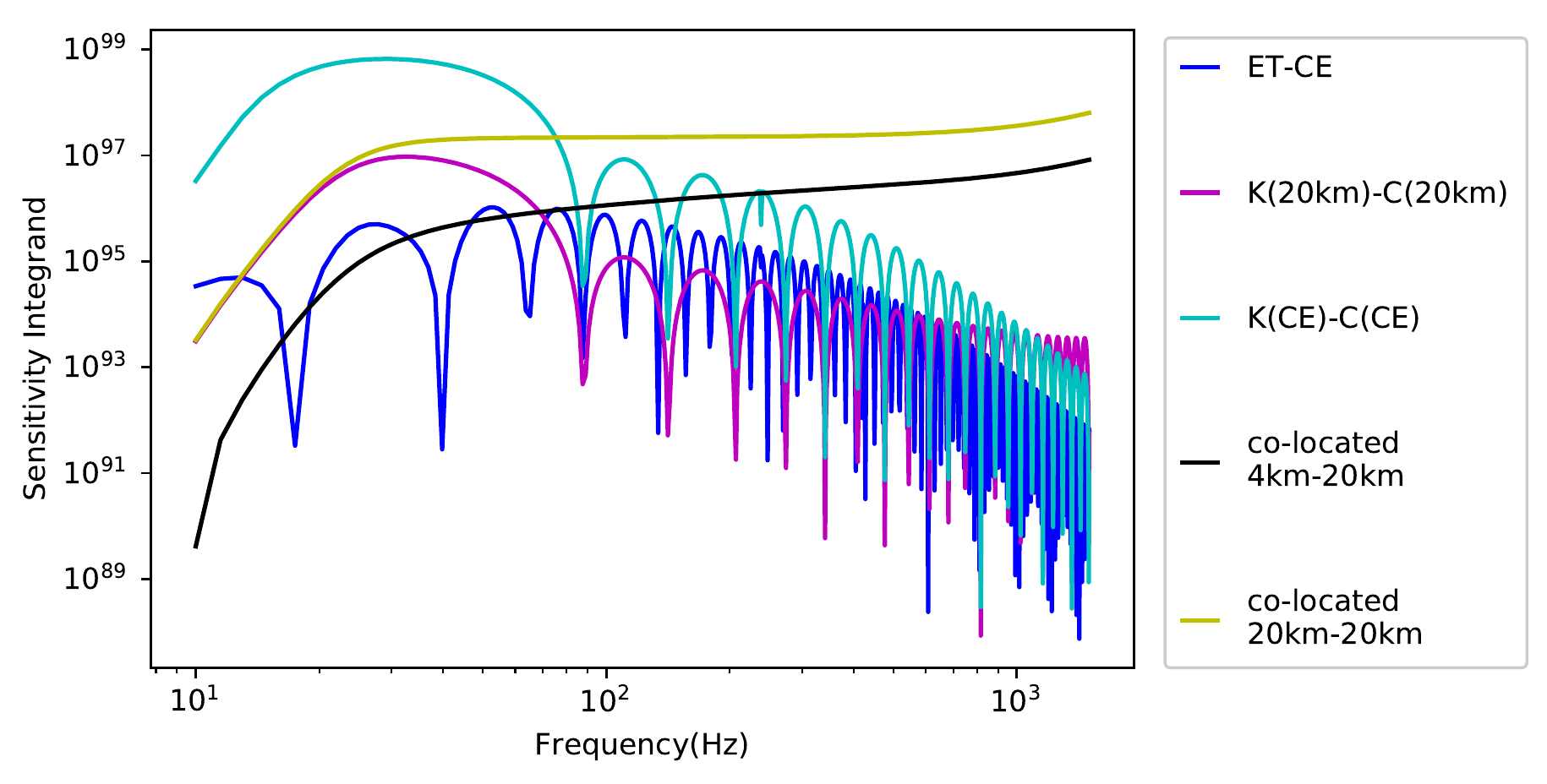}
\end{minipage}
\begin{minipage}[t]{0.5\textwidth}\
\includegraphics[width=8.6cm]{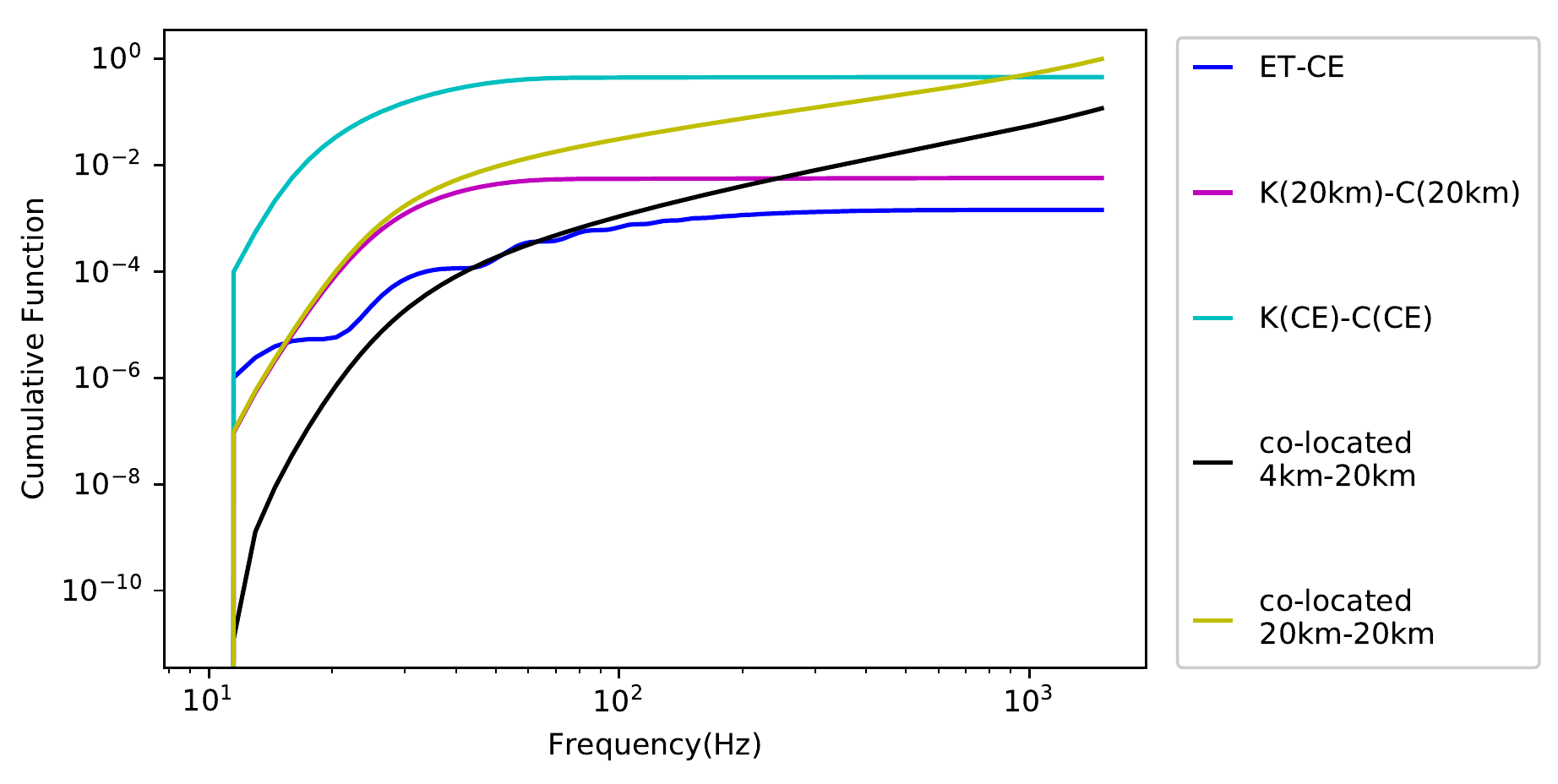}
\end{minipage}
\caption{Left panel: the sensitivity integrand versus frequency plot. The sensitivity integrand illustrates the co-response of gravitational wave detector network to the sensitivity of different frequencies. The area under every curve, respectively, represents the detectability of every detector network configuration. The larger the area under the curve, the more stronger constraining power of the corresponding detector network, meaning this detector network configuration corresponds to a lower line in Figure~\ref{res}. If the overlap reduction function is optimal, namely, the detectors are co-located and have the same arm orientation, the curve will level off when it tends to high frequencies, for example, `co-located 4km-20km' detectors in black and `co-located 20km-20km' detectors in yellow. And  if the overlap reduction function is not optimal, the curve will begin oscillating down rapidly at certain frequency, which is generally less than 100Hz, for example,  `ET-CE' in blue, `K(20km)-C(20km)' in purple, and `K(CE)-C(CE)' in sky blue. Although the curve of `K(CE)-C(CE)' begins oscillating rapidly approximately at 100Hz, the sensitivity integrand in frequency range of [10,100]Hz is large enough to make this configuration have an excellent performance for stochastic background detection in string cosmology, especially in frequency band of lower than 1000Hz. Right panel: the vertical axis is the ratio of the cumulative function of the sensitivity integrand of multiple detector pairs to the integral of the sensitivity integrand over frequency of `co-located 20km-20km' case, because it's value is the largest in all five detector pairs. The horizontal axis is frequency. The piecewise frequency contribution for S/N can be seen more clearly from this plot. It can be concluded from this plot that the main contribution of nonoptimal overlap reduction function case for stochastic background detection in string cosmology comes from low-frequency parts, for example, for `ET-CE', `K(20km)-C(20km)' and `K(CE)-C(CE)', the cumulative function is basically unchanged after 100Hz, while for optimal overlap reduction function case, the main contribution for stochastic background detection in string cosmology comes from high-frequency parts, for example, for `co-located 4km-20km' and `co-located 20km-20km', the cumulative function has maintained growth in the frequency range of this plot. It means that for high-frequency truncated stochastic background in string cosmology, the colocated detectors have better performance than separate detectors. }\label{SI}
\end{figure*}
In order to present the contribution of different frequency bands to S/N more clearly, we calculated the `sensitivity integrand'\citep{2007CQGra..24S.639C} for multiple detector pairs in Figure~\ref{SI}, which allows for a deeper understanding of the previous results. The `sensitivity integrand' illustrates the contribution to the sensitivity of different frequencies:
\begin{equation}
\mathcal{I}_{12}(f)=\frac{\left[\gamma_{12}(f)\right]^{2}}{P_{1}(f) P_{2}(f)}.
\end{equation}
Combining Eq.~\ref{SN1} with Eq.~\ref{dsp} and Eq.~\ref{do1}, it can be easily found that the integral of the sensitivity integrand is an intuitive quantity to represent S/N. For a given model, the larger the integral of the sensitivity integrand, the larger of the S/N. And for a given S/N, the larger the integral of the sensitivity integrand, the better of the constraining power of detector networks for stochastic background detection; this will correspond to a lower line in Figure~\ref{res}. The sensitivity integrand of multiple detector pairs in the 3G case has been presented in the left panel of Figure~\ref{SI}, in which the vertical axis is the sensitivity integrand, and the horizontal axis is frequency. The area under every curve, respectively, represents the detectability of every detector network configuration for stochastic background detection. We can conclude from Figure~\ref{SI} that if the overlap reduction function is optimal, namely, the detectors are colocated with the same arm orientation, the curve will level off when it tends to high frequencies, for example, `colocated 4km-20km' detectors in black and `colocated 20km-20km' detectors in yellow. If the overlap reduction function is not optimal, the curve will begin oscillating down rapidly at a certain frequency, which is generally less than 100Hz, for example,  `ET-CE' in blue, `K(20km)-C(20km)' in purple, and `K(CE)-C(CE)' in sky blue. Although the curve of `K(CE)-C(CE)' begins oscillating rapidly approximately at 100Hz, the sensitivity integrand in the frequency range of [10,100]Hz is large enough to make this configuration have the best performance for stochastic background detection in string cosmology.  The main contribution of the nonoptimal overlap reduction function case for stochastic background detection in string cosmology comes from low-frequency parts, while for the optimal overlap reduction function case, the main contribution for stochastic background detection in string cosmology comes from high-frequency parts. This means that for high-frequency truncated stochastic background in string cosmology, the colocated high-frequency detectors have better performance than separate detectors. Furthermore, in the right panel of Figure~\ref{SI}, the vertical axis is the ratio of the cumulative function of the sensitivity integrand of multiple detector pairs to the integral of the sensitivity integrand over frequency of the `colocated 20km-20km' case, because this value is the largest in all five detector pairs, and the horizontal axis represents frequency. The piecewise frequency contribution for signal-to-noise ratio can be seen more clearly from this plot. It is more evident from this plot that the main contribution of the nonoptimal overlap reduction function case for stochastic background detection in string cosmology comes from low-frequency parts, for example, for `ET-CE', `K(20km)-C(20km)' and `K(CE)-C(CE)', the cumulative function is basically unchanged after 100Hz, while for the optimal overlap reduction function case, the main contribution for stochastic background detection in string cosmology comes from high-frequency parts, for example, for `colocated 4km-20km' and `colocated 20km-20km', the cumulative function has maintained growth in the frequency range of this plot. This further confirms that if the stochastic gravitational wave background spectrum is high frequency truncated, then colocated detectors have great  advantages for stochastic background detection.

\section{Conclusion}\label{sec:highlight}
In this work, we discussed how to use the detector's information, i.e., noise power spectral density, location, and orientation, to give an estimation of stochastic gravitational wave background with at least a 95\% detection rate and a 5\% false alarm rate. Furthermore, how to improve the performance of the detector network by the sensitivity curve and location together with the arm orientation of the detector has been discussed. We have found that the performance of the detector network for stochastic background will be optimal with higher sensitivity and larger overlap reduction function.  The overlap reduction function will be optimal if  detectors are close enough with parallel arm orientation, and the angle between the line joining detectors and their arm orientation also affects the overlap reduction function: a smaller angle leads to a larger overlap reduction function. Based on this information, we select a location in Wuhan in China, to evaluate its contribution to a future GWD network. \\
\indent We can draw a conclusion from Figure~\ref{res} in the results section that CKGO together with KAGRA can achieve a better overlap reduction function than current LIGO detector pairs, and therefore lead to a better constraining power for stochastic background detection. Furthermore, with an optimal overlap reduction function, a colocated detector will lead to excellent performance in the high-frequency range for stochastic background detection. In the 2G era, CKGO-KAGRA can give a more stringent limit for stochastic background than LIGO detector pairs in a frequency range of higher than~50Hz. As long as KAGRA-CKGO exists in the network, the whole performance will be better than any other configuration. If one of the LIGO detector pairs is offline, KAGRA-CKGO can work as an alternative to LIGO detector pairs. In the future 2.5G era, when LIGO will upgrade to LIGO voyager and China is planning to build a 4km baseline detector, if we apply the sensitivity curve of 4km detector to KAGRA and CKGO, and apply the sensitivity curve of LIGO voyager to LIGO detector pairs, we will find that the LIGO detector pairs have better performance than KAGRA-CKGO for stochastic background detection. This is mainly due to the fact that the sensitivity curve applied to LIGO pairs~(LIGO voyager) is better than that of KAGRA and CKGO~(planned 4km detector) below a frequency of 600Hz. Furthermore, if the sensitivity curve of LIGO voyager is applied to KAGRA and CKGO, the results will turn out to be that KAGRA-CKGO have better performance than LIGO detector pairs. In the 3G era, apparently, KAGRA-CKGO will apply a  sensitivity curve of CE or 20km, both configurations will have better performance than ET-CE. In Figure~\ref{SI}, we can draw a conclusion that the main contribution of the nonoptimal overlap reduction function case for stochastic background detection in string cosmology comes from low-frequency parts, while for the optimal overlap reduction function case, the main contribution for stochastic background detection in string cosmology comes from high-frequency parts. Thus with the optimal overlap reduction function, high-frequency colocated detectors have excellent performance for stochastic background detection in string cosmology when it tends to higher frequency. This means that for a high-frequency truncated stochastic background in string cosmology, the colocated high-frequency detectors have better performance than separate detectors. Overall, establishing a detector in China and colocated detectors can make a remarkable contribution to the detector network for stochastic background detection, and CKGO together with KAGRA can work as a promising alternative when one of the LIGO detector pairs is offline in the future. For the colocated case, high-frequency detectors could play an important role in exploring the stochastic gravitational wave background in string cosmology in a frequency range of higher than 1000Hz. Thus, taken together with building a GWD in China, the colocated high-frequency detector pairs would be very profitable for stochastic background detection. The earlier the CGWD is built, the sooner the detection of the stochastic gravitational wave background will benefit from it. \\
\indent This paper provides instructions on how to select the location and arm orientation of a detector for better stochastic background detection. Gravitational wave astronomy has developed faster and faster as detectors have been upgraded. In the current O3 run, started on 2019 April 1st, an average of one gravitational wave event is detected per week. Gravitational wave events are becoming routine events, thus it is essential to establish the pipeline from theory to real detection, so as to get some answers about the universe from the large sample of events. In this sense, the discussion in this paper is meaningful for the future location selection of GWDs and constraints of proper models.  \\

\acknowledgments
We thank HaiXing Miao for providing the sensitivity curve of 4km and 20km detector and valuable discussions. We are also grateful to R.Y.L, Z. Liao and S. Yu for helpful discussions.  Y.L. and L.G. are supported by the National Program on Key Research and Development Project through grant No. 2016YFA0400804, and by the National Natural Science Foundation of China with grant No. Y913041V01, and by the Strategic Priority Research Program of the Chinese Academy of Sciences through grant No. XDB23040100. X. F. is supported by the National Natural Science Foundation of China(under Grants No.11673008, 11633001 and 11922303), the Strategic Priority Program of the Chinese Academy of Sciences (Grant No. XDB 23040100) and Newton International Fellowship Alumni Follow-on Funding.

\end{document}